\documentclass[aps,showpacs,epsf,twocolumn,prl]{revtex4}

\usepackage{amsmath}
\usepackage{amssymb}
\usepackage{bbold}
\usepackage{mathrsfs}
\usepackage{graphicx, psfrag}
\usepackage[centerlast]{caption}
\usepackage{float}
\usepackage{subfig}
\usepackage{tikz}
\usepackage{pgfplots}
\usepackage[colorlinks=true, citecolor=blue, urlcolor = blue, linkcolor= red, bookmarks=true]{hyperref}
\usepackage{epstopdf}

\begin{document}
\def \beq{\begin{equation}}
\def \eeq{\end{equation}}
\def \bse{\begin{subequations}}
\def \ese{\end{subequations}}
\def \bea{\begin{eqnarray}}
\def \eea{\end{eqnarray}}
\def \bem{\begin{displaymath}}
\def \eem{\end{displaymath}}
\def \bem{\begin{pmatrix}}
\def \eem{\end{pmatrix}}
\def \beb{\begin{bmatrix}}
\def \eeb{\end{bmatrix}}
\def \bc{\begin{center}}
\def \ec{\end{center}}
\def \bb{\bibitem}
\def \bs{\boldsymbol}
\def \nn{\nonumber}
\def \mf{\tilde{J}_1}
\def \mj{\tilde{J}_0}
\def \mh{\mathcal{H}}
\def \ma{\mathcal{A}}
\def \md{\mathcal{D}}
\def \mg{\mathcal{G}}
\def \hkx{\hat{k}_{x}}
\def \hky{\hat{k}_{y}}
\def \bq{\bar{q_{y}}}
\newcommand{\cc}{\color{red}}
\newcommand{\cb}{\color{blue}}
\newcommand{\upa}{\uparrow}
\newcommand{\dna}{\downarrow}
\newcommand{\pdag}{\phantom\dagger}
\newcommand{\bra}[1]{\langle #1 |}
\newcommand{\ket}[1]{| #1 \rangle}
\newcommand{\braket}[2]{\langle #1 | #2 \rangle}
\newcommand{\ketbra}[2]{| #1 \rangle \langle #2 |}
\newcommand{\expect}[1]{\langle #1 \rangle}
\newcommand{\inx}[1]{\int d\bs x \Big [ #1 \Big ]}
\newcommand{\si}[1]{\Psi^{\pdag}_{#1} (\bs x)}
\newcommand{\dsi}[1]{\Psi^\dag_{#1} (\bs x)}
\newcommand{\evolve}[1]{\frac{\partial #1}{\partial t}}

\title{Entanglement like properties in Spin-Orbit Coupled Ultra Cold Atom and violation of Bell like Inequality}
\author{ Rahul Kumar and Sankalpa Ghosh}
\affiliation{Department of Physics, Indian Institute of Technology Delhi, New Delhi-110016, India}

\begin{abstract}
We show that the general quantum state of synthetically spin-orbit coupled ultra cold bosonic atom whose condensate was  experimentally created recently ( Y. J. Lin {\it et al.}, 
Nature, {\bf 471}, 83, (2011)), shows entanglement between motional degrees of freedom ( momentum) and internal degrees of freedom (hyperfine spin). We demonstrate the violation of Bell-like inequality (CHSH) for  such states  that provides a unique opportunity to verify fundamental principle like quantum non-contextuality for commutating observables  which are not spatially separated. We analyze in detail the Rabi oscillation executed by such atom-laser system and how that influneces quantities like entanglement entropy, violation of Bell like Inequality etc. We also discuss the implication of our result in testing the quantum non-contextuality and Bell's Inequality vioaltion by macroscopic quantum object like Bose-Einstein Condensate of ultra cold atoms. 
\end{abstract}
\maketitle
\section{Introduction}

Entanglement,  a highly enigmatic feature of quantum mechanics implies a non-local correlation between two ( or more ) quantum systems such that the description of their states has to be done in reference to the each other even when they are separated by space like interval. This is a direct consequence of the fact that a state of composite quantum system can be expressed as a linear supersposition of tensor product of the states corresponding to its different parts. This was first pointed out by
E. Schrodinger \cite{sch} and around the same time by Einstein, Podolsky and Rosen (EPR) \cite{EPR} who 
cited this property to question the compatibility of quantum mechanics with local realism. 
Almost a half century back in a milestone paper, 
J. S. Bell \cite{Bell} addressed this fundamental question raised by EPR  with the help of  Bell's Inequality(BI)
whose violation denies the existence of Local Hidden variable (LHV) theories and  
validates the existence of EPR like non-local correlation in quantum mechanical states.  The demonstration of 
the violation of BI in experiments \cite{Aspect} thus forms a fundamental test to validate quantum mechanics as the the right physical theory.

Subsequent works indicates that the BI has more general implication like quantum non-contextuality \cite{Bell1, KS, Mermin} which means that the outcome of a particular measurement is determined independently of previous (or simultaneous) measurement of any set of mutually commutating observables \cite{Roy, Sayandeb}. Only when such set of observables commutes due to their space like separation, they correspond to the cases for which EPR like non-local correlation is relevant.  A true test of quantum non-contextuality thus requires exploration of entanglement like properties and the test for BI in commutating observables that are not spatially separated. Availability of such experimental systems is not very common\cite{Yuji}.

In this work we show that synthetically spin-orbit coupled (SSOC) ultra cold bosonic atoms that was recently 
realized in experiments \cite{Galitski, Spielman}
provides a quatum system where bipartite entanglement like quantum correlation 
occurs between motional observables ( momentum) and intrinsic observable ( spin) of a given atom and thus provides us an unique opportunity to test the concept of quantum non-contextuality for commutative observables of same 
object, that are not spatially separated. Ultra cold atomic systems \cite{Book1, Book2, review1, Book3, review2} are very clean, almost bereft of any thermal fluctuation and highly isolated from 
the classical envirornment and thereby form one of the most ideal quantum systems. 
Spin-orbit coupled ultra cold atomic system \cite{Galitski, Spielman, FermiSO1, FermiSO2, review3, review4} is one of most significant development in recent times in this direction and particularly being explored extensively for the possibility of quantum simulating novel topological condensed matter phases \cite{Topo} with 
such spin-orbit coupled ultra cold atoms. 

In this context our work extend the outreach of such ultra cold quantum system in a different direction 
by showing that it has the potential of testifying fundamental question associated with the quantum world such as non-contextuality. 
Additionally the experiment on SSOC Bose Einstein Condensate (BEC) \cite{Spielman} already suggest that even though the system is interacting, its phase diagram can be understood from a non-interacting single-boson effective hamiltonian. 
Therefore our results rigorously derived for spin-orbit coupled ultra cold bosonic atom 
may also be applied for SSOC BEC 
which is a macroscopic quantum objects consists of a large number of atoms in quantum mechanical ground state. This provides an opportunity of studying quantum non contextuality, BI violation, entanglement like properties 
for a macroscopic and massive quantum objects, an instance which is not commonly available \cite{massive,
macro}.  

Using well established definitions for entangled states we first demonstrate that the general quantum mechanical states of such SSOC ultra cold bosons satisfy the criterion of bipartitite entanglement and we quantify such entanglement with well known entanglement measure like 
entanglement entropy and concurrence. We particularly show how the  dressed states of such atoms  execute a Rabi like oscillation between two different energy minimums in the momentum space and how such oscillation leads to the similar oscillation in well known entanglement measure like entanglement entropy and concurrence. 
Finally we show how such entanglement between motional and spin-degrees of freedom can lead to the BI violation. 
\begin{figure}
\centering
\includegraphics[width=0.95\columnwidth , height= 
0.75\columnwidth]{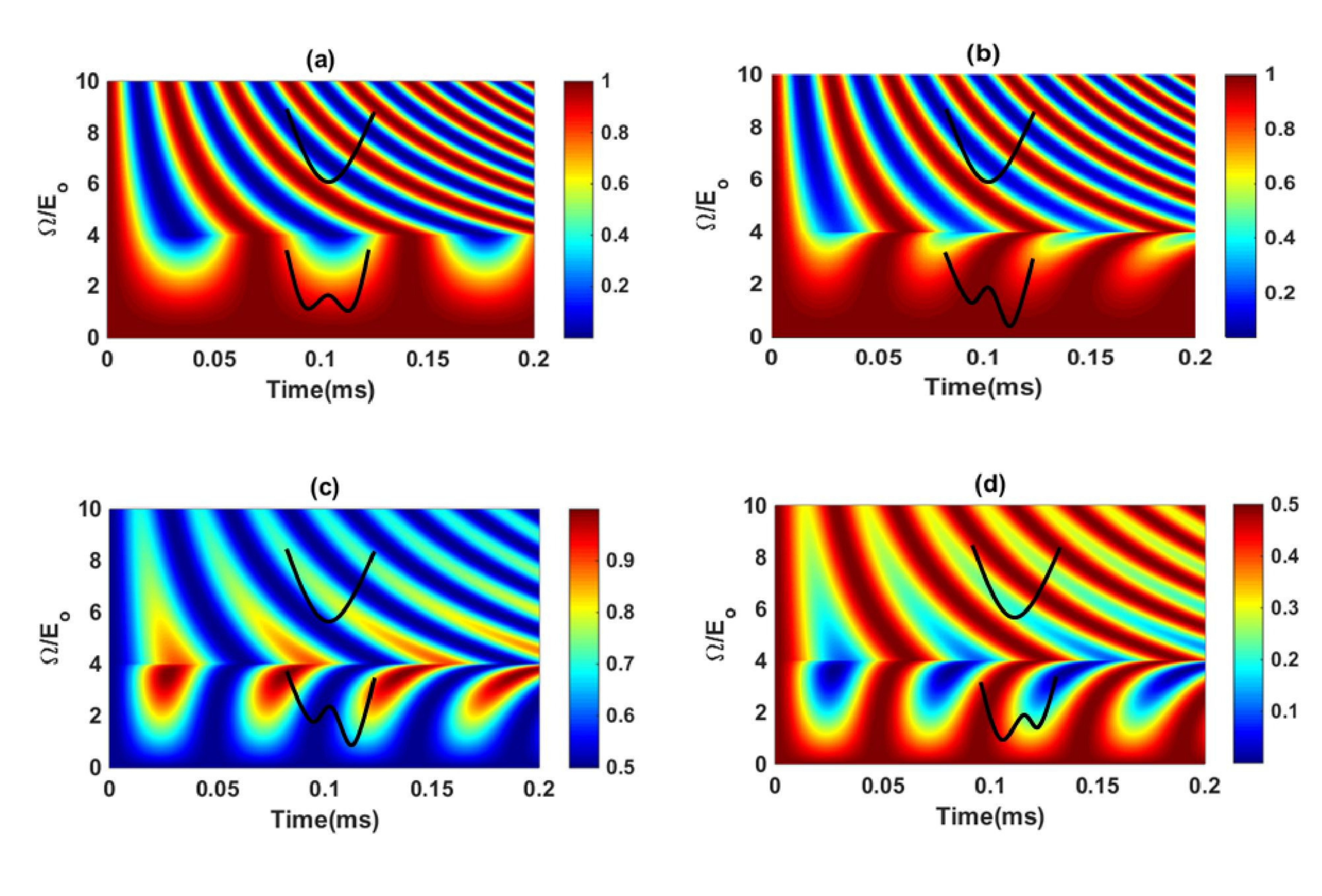} 
\caption{(Color online) The color axis in each figure plotted $|C_{2}(t)|^{2}$. 
Fig. (a) and (b) with the initial condition $C_{1}=0, C_{2}=1$ with $\delta=0, 2E_{0}$ respectively. Fig. (c) and 
(d) are with initial condition $C_{1}=C_{2}=\frac{1}{\sqrt{2}}$ and $\delta = \pm 2E_{0}$ respectively. And for figures $\lambda_{0}=\frac{2\pi}{k_{0}}=804.1nm$.The black lines depict the ground state energy dispersion (see the text
for details).}
\label{fig:Rabi}
\end{figure}
\section{Basic Formalism: Hamiltonian and Time Evolution}
The SSOC system considered here was experimentally realized in NIST experiment \cite{Spielman}  and consists of ultra cold $^{87}$Rb atoms in two internal states, $| m_{F} \rangle = |0\rangle, | -1 \rangle$  available in $F=1$ hyperfine state of $5S^{\frac{1}{2}}$ electronic levels. 
A pair of counterpropagating Raman lasers couple these states and the non-interacting/single-atom hamiltonian for the corresponding system can be written as 
$H_{SO}' = \frac{\hbar^{2}(k_{x}^{2}+ k_{0}^{2})}{2m} I + \frac{\delta}{2} \sigma_{y} + \frac{\Omega}{2}\sigma_{z} + \frac{2 \hbar^{2}k_{0}k_{x}}{2m} \sigma_{y}$.
After performing a  suitable pseudospin rotation on the above hamiltonian \cite{Hui}
 it can be rewritten as 
\beq
H_{SO}=\begin{bmatrix}
\frac{\hbar^2}{2m}(k_x+k_o)^2+\frac{\delta}{2} & \frac{\Omega}{2}\\
\frac{\Omega}{2} & \frac{\hbar^2}{2m}(k_x-k_o)^2-\frac{\delta}{2}
\end{bmatrix}. \label{HSO}
\eeq Here $k_{x}$ and $k_{0}$ are repectively atom and photon momentum, $\Omega$ is the Rabi coupling strength and $\delta=\hbar (\Delta \omega_{L} - \omega_{z})$, where  $\Delta \omega_{L}$ is the detuning between the Raman lasers and $\omega_{z}$ is the Zeeman frequency.
Since  the properties of the ultra cold BEC of SSOC bosons can be understood through this non-interacting Hamiltonian (\ref{HSO}) \cite{Spielman} our  subsequent discussion will be based on the hamiltonian. 
Diagonalization of (\ref{HSO}) gives energy eigenvalues  
$E_{g,e} = E \mp A$
and the energy eigenstates as 
\bea |g \rangle & = & C_{+}| \uparrow, k_{x} + k_{o} \rangle + C_{-}| \downarrow, k_{x} - k_{o} \rangle \nn \\
|e \rangle & = & -C_{-} | \uparrow, k_{x} + k_{o} \rangle + C_{+} | \downarrow, k_{x} - k_{o} \rangle \label{dress} \eea
with $C_{\pm}=\frac{1}{\sqrt{2A}}( A \pm \sqrt{A^{2} - \frac{\Omega^{2}}{4}})^{\frac{1}{2}}$.
%
Here the suffix $_{g}, _{e}$ respectively denote ground  and excited state of the dressed atom ( dressed states), 
$E_{x}= \frac{\hbar^{2}k_{x}^{2}}{2m}$ and $E_{0}= \frac{\hbar^{2}k_{0}^{2}}{2m}$, such that

\bea E &=& E_{x} + E_{0} \nn \\
   A & = & 2 E_{0} \sqrt{ \frac{E_{x}}{E_{0}}( 1 + \frac{\delta}{4 \sqrt{E_{x}E_{0}}})^{2} + (\frac{\Omega}{4E_{0}})^{2}} \label{param} \eea
and the  states $|\downarrow,k_x-k_o\rangle$ and $|\uparrow,k_x+k_o\rangle$ are the bare spin-orbit eigenstates.
It may be pointed out that even if the interaction is included in the mean field approximation, the condensate wave-function can be expressed as a linear combination of these bare spin-orbit eigenstates \cite{Zheng}. 
Since the atoms form a BEC \cite{Spielman} , $k_{x}$ in expression (\ref{param}) are determined from the minimum energy condition. 
 The dispersion $E_{g}(k_{x})$ has respectively single and double well like structure  for $\Omega \ge 4E_{0}$ and  $\Omega < 4E_{0}$ respectively with the corresponding minimas at $k_{x}=0$ in the first case and at $k_{x}=\pm k_{0}\sqrt{1-\Big(\frac{\Omega}{4E_{0}}\Big)^2}$ for the double well.  These dispersion is shown schematically ( black line) in  for $\delta =0$ in Fig. \ref{fig:Rabi}
(a) and for $\delta \neq 0$  in Fig. \ref{fig:Rabi} (b), (c) and (d).
The time evolution under (\ref{HSO}) is given by    
$ i \hbar \frac{d |\Psi \rangle}{dt} = H_{SO} | \Psi \rangle $
and the general time dependent normalized state can be written as 
\bea | \Psi_{SO} (t) \rangle & = & a \exp ( -\frac{i E_{g} t}{\hbar}) | g \rangle + b \exp ( - \frac{i E_{e} t}{\hbar}) |e \rangle \nn \\
& = & C_{1}(t) | \downarrow, k_{x} - k_{o} \rangle   + C_{2}(t)| \uparrow, k_{x} + k_{o} \rangle  \label{ENT} \eea 
with ( $\sigma_{z}, \sigma_{x}$ are Paul matrices)
\beq \begin{bmatrix} C_{1}(t) \\ C_{2}(t) \end{bmatrix} = e^{-i\frac{E_{0} t}{\hbar}}[b\exp (\frac{iAt}{\hbar}))\sigma_{z} + a\exp ( \frac{-iAt}{\hbar}))\sigma_{x}]\begin{bmatrix} C_{+} \\ C_{-} \end{bmatrix} \eeq
to be determined from the initial condition.  
For example, if the atom is prepared in pseudospin (up)-momentum state $|\uparrow, k_{x}+k_{0} \rangle$. Then $|\Psi_{SO}(0) \rangle = | \uparrow, k_{x} + k_{o} \rangle \Rightarrow C_{1}(0) =0, C_{2}(0)=1$
which gives 
$a= C_{+} ; \quad b = -C_{-}$. 
The corresponding time evolved state at $t \ge 0$ becomes
\bea 
|\Psi_{SO}(t) \rangle & =& [\cos \frac{\pi t}{T} - i\frac{\sqrt{A^{2} - \frac{\Omega^{2}}{4}}}{A}\sin \frac{\pi t}{T}]e^{-i\frac{E_{0} t}{\hbar}}  | \uparrow, k_{x} + k_{o} \rangle \nn \\
& & \mbox{} -i \frac{\Omega}{2A} [ \sin \frac{\pi t}{T}]e^{-i\frac{E_{0} t}{\hbar}} | \downarrow, k_{x} - k_{o} \rangle 
\label{Rabi}\eea
The probability $|C_{2}(t)|^{2}$ is plotted as a function of $t$ and  $\frac{\Omega}{E_{0}}$
with this initial condition in Fig. \ref{fig:Rabi} (a) and(b) respectively for zero and finite $\delta$. The change in the ground state energy dispersion above and below $\frac{\Omega}{E_{0}}=4$ is also indicated schematically in each plot. From (\ref{Rabi}) the time period for oscillation for $|C_{1,2}(t)|^{2}$
is $T=\frac{\pi\hbar}{A}$. 
At  $t=nT, n \in \mathcal{Z}$  with 
$|\Psi_{SO}(nT) \rangle=\cos(n\pi)e^{\frac{-iE_onT}{\hbar}}|\uparrow,k_x+k_o\rangle \Rightarrow |C_{2}(t)|^{2}=1$.  
For $nT \le t \le (n+1)T$, the amplitude get transferred to 
$|\downarrow, k_{x}-k_{0} \rangle$ state.    

Through (\ref{param})  $A$, inverse of the time period $T$, depends on $E_{x}$ and $\frac{\Omega}{E_{0}}$
for a given value of $\delta$.
Now, $k_{x}$ and $E_{x}(k_{x})$ changes abruptly at $\Omega=4E_{0}$ .  Consequently when 
the value of  $\frac{\Omega}{E_{0}}$ is continuously decreased, the time period $T$ also changes 
continuously. But at the point $\Omega=4E_{0}$, there is an abrupt change in $T$ 
as can be in Fig. \ref{fig:Rabi} (a) and(b). When the coupling $\Omega \sim 0$, the system stays almost always in the initial state with $C_{2}=1, C_{1}=0$ as can be seen in Fig. \ref{fig:Rabi} (a) and(b). The  transition probability to $|\downarrow, k_{x}-k_{0} \rangle$, $|C_{1}(t)|^{2}$ increases with increasing $\Omega$. It also depends on whether the detuning $\delta$ is zero, positive or negative. The  Fig. \ref{fig:Rabi} (c) and (d) also plots $|C_{2}(t)|^{2}$ 
but with initial condition $C_{1}(0)=C_{2}(0)$  respectively with $\delta > 0$ and $\delta < 0$. 
\begin{figure}
\centering
\includegraphics[width=0.95\columnwidth , height=  
0.65\columnwidth]{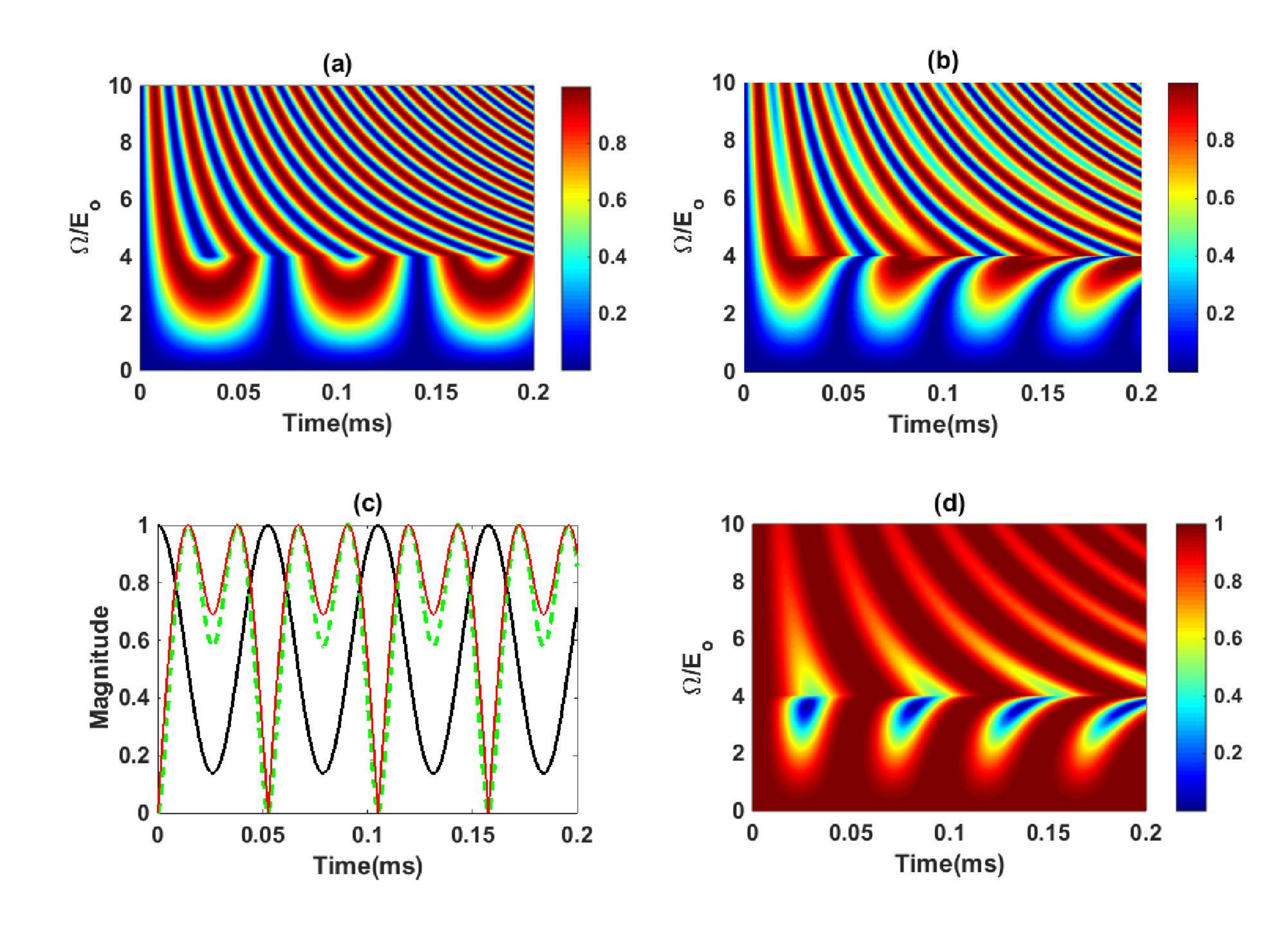} 
\caption{(Color online) Fig. (a), (b) plot the entanglement entropy (S) along the color axis as a function of time 
and $\frac{\Omega}{E_{0}}$ for initial condition $C_{1}=0, C_{2}=1$ with all other conditions same as 
Fig. \ref{fig:Rabi}(a) and (b). Fig. (d) is a similar plot but with 
 initial condition $C_{1}=C_{2}=\frac{1}{\sqrt{2}}$ and other conditions same as Fig. (b).  Fig. (c) gives a comparative
plot for the time evolution of $|C_{2}(t)|^{2}$ (black thick solid line), $S_{\epsilon}$ (green dashed line) and concurrence C 
( red thin solid line) for initial condition $C_{1}=0, C_{2}=1$.}
\label{fig:entropy}
\end{figure}
\section{Entanglement and its Measure}
 The desnity matrix for the state (\ref{ENT})  
\beq
\rho_{SO}=| \Psi\rangle_{SO}\langle\Psi |_{SO} =
\begin{bmatrix}
|C_1(t)|^2 & 0 & 0& C_1(t)C_2(t)^*\\
0& 0& 0& 0\\
0& 0& 0& 0\\
C_1(t)^*C_2(t)&0& 0 & |C_2(t)|^2\\
\end{bmatrix} \label{dmso}
 \eeq
The matrix elements of (\ref{dmso}) satisfies the condition  $\rho_{ii}\rho_{jj}=\rho_{ij}\rho_{ji}$, criterion for a pure state. Now we investigate the entanglement like properties for this state (\ref{ENT}) and show that it satisfies the criterion  for an bipartite entangled state where the entanglement is now not between spatially separated two subsystem for  same degrees of freedom, but rather between two different degrees of freedom of the same system. The basis states in the pseudospin space $H_{s}$ and the momentum(orbital) space $H_{k}$
\beq |u_{1} \rangle  =  | \downarrow \rangle,  |u_{2} \rangle =| \uparrow \rangle, |v_{1,2} \rangle  =  |k_{x} \mp k_{0} \rangle \label{basis} \eeq
A general state in $H_{s}$ and in $H_{k}$ can be written as, 
$ |\Psi\rangle_S  =  a|\uparrow\rangle  +b|\downarrow\rangle, |\Psi\rangle_m  =  c|k_x-k_o\rangle  + d|k_x+k_o\rangle $
It follows directly that for $a,b,c,d \ne 0$ that 
\beq | \Psi_{SO}(t) \rangle 
\neq |\Psi\rangle_S \otimes |\Psi\rangle_m \eeq
The above conclusion for the  state (\ref{ENT}) can also be verified from the Schimdt decompositin criterion of 
a pure state that allows one to write it as an expansion of biorthogonal terms $|u_{i} \rangle \otimes |v_{i} \rangle$ \cite{Ekert} ($i=1,2$) so that  
$ | \Psi_{SO} \rangle = \sum_{i} g_{i} | u_{i} \rangle \otimes | v_{i} \rangle$
with $\sum_{i} |g_{i}|^{2} =1$. Here $g_{i}$'s are called Schimdt coefficient and number of such $g_{i}$ are called the Schimdt rank. The crterion for bipartite entanglement is that is Schimdt rank is $ > 1$. Now from (\ref{basis}) we can directly check that the Schimdt rank for the state $|\Psi_{so} \rangle$ is $2$. Therefore the state satisfies the property of an entangled state.  
Also from (\ref{dmso}), the reduced density matrix for (pseudo)spin and momentum are respectively
\beq \rho_{s}= Tr_m\rho_{so} =  \begin{bmatrix} 
|C_1(t)|^2 & 0\\
0 & |C_2(t)|^2\\
\end{bmatrix} = \rho_{m}=Tr_s\rho_{so}  
\label{dmsk}\eeq
From (\ref{dmso}) and (\ref{dmsk})
clearly $\rho_{SO}\neq \rho_s\otimes\rho_m$  and this also confirms the entanglement like behavior in the system. 
The entanglement of the state in Eq. \ref{ENT} can also be verified through 
PPT(Positive Partial Transposition) criterion given by Peres and Horodecki \cite{Peres,Horo}
which is a necessary and sufficient condition for separability for  
dim$(H_1\otimes H_2)\leq 6$). For the present case  $1$ and $2$ referes to spin and orbit(momentum). 
and the dimension of the product space  is $4$. 
After partial transposition (block-wise transposition) of matrix $\rho_{SO}$ given in (\ref{dmso}) becomes 
\beq
\rho_{SO}^T=\begin{bmatrix}
|C_1(t)|^2 & 0 & 0& 0 \\
0& 0& C_1(t)C_2(t)^*& 0\\
0& C_1(t)^*C_2(t)& 0& 0\\
0&0& 0 & |C_2(t)|^2\\
\end{bmatrix}  \label{PPTDM} \eeq
The eigenvalues of (\ref{PPTDM}) are given by 
$\sqrt{|C_1(t)|^2|C_2(t)|^2},-\sqrt{|C_1(t)|^2|C_2(t)|^2},|C_1(t)|^2,|C_2(t)|^2$. 
Since $\rho_{so}^T$ has atleast one negative eigenvalue, therefore acording to the PPT criterion it has  entanglement between spin and momentum degree of freedom.

\begin{figure}
\centering
\includegraphics[width=1.05\columnwidth , height= 
0.45\columnwidth]{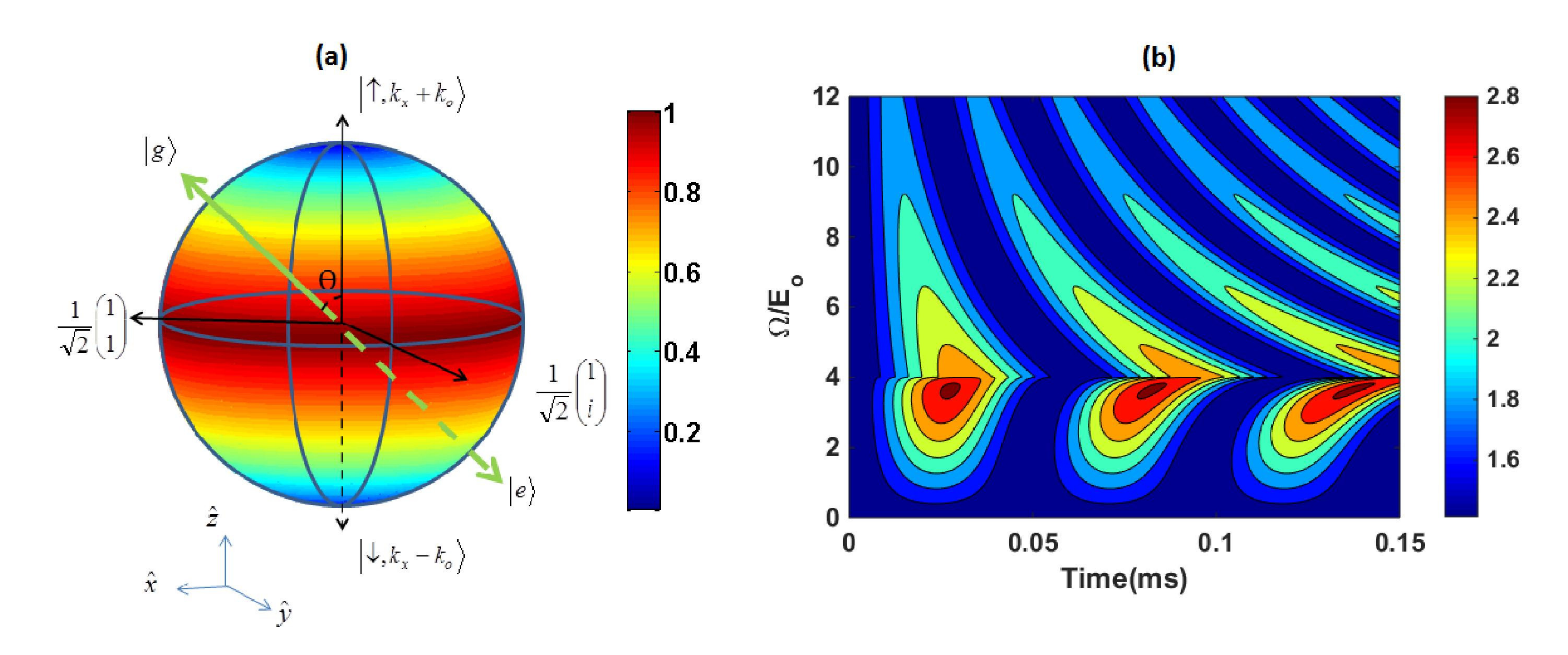} 
\caption{(Color online) (a) The dressed state of the atom (pure state) defined in (\ref{dress})
on the surface of the Bloch sphere  with varying $\Omega$. The color gives the value of entanglement entropy $S_{\epsilon}$ defined in (\ref{EE}).  For all such states azimuthal angle $\phi=0$ 
and polar angle $\theta=\arctan ( \frac{\frac{\Omega}{4E_{0}}}{\sqrt{\frac{E_{k_{x}}}{E_{0}}}+ \frac{\delta}{4E_{0}}})$
and such dressed states lies in the $x-z$ plane. The north and south pole corresponds to bare states which are product states and the maximally entangled state or Bell state lies in the equilateral plane.
(b) The Bell coefficient ${\ss}$ ( color-axis) as a function of time ($x$-axis) and $\frac{\Omega}{E_{0}}$ ($y$-axis) for conditions corresponding to Fig. \ref{fig:Rabi}(a).}
\label{fig:BB}
\end{figure}

Once we demonstrate the entanglement in (\ref{ENT}), we can now try to quantify this entanglement for SSOC ultra cold bosons using a suitable measure for entanglement \cite{measure1} such as the von Neuman entropy (deonoted as $S$). It is known that for any given pure entangled state $S=0$, whereas the entropy of the subsystems is greater than zero \cite{quantum} since the state corresponding to each subsystem of this bipartite system is a mixed state.  

The above cirterion for the von Neuman entropy can be established for the system under consideration. By definition \cite{quantum} 
$S(\rho) = -Tr(\rho\log \rho)= -\sum_i \lambda_i\log\lambda_i $
where $\lambda_i$ are eigenvalue of the density matrix $\rho$.
For the density matrix given in (\ref{dmso}) the eigenvalues are respectively $1,0,0,0$. 
which readily gives $S(\rho)=0$ as expected. 
Similarly from the density matrices in  (\ref{dmsk}),  $S$ for the subsystems can be determined as 
$S(\rho_{s})=S(\rho_{m})=S_{\epsilon}$ with  
\beq
S_{\epsilon} = -|C_1(t)|^2\log|C_1(t)|^2-|C_2(t)|^2)\log|C_2(t)|^2  \label{EE} \eeq 
and satisfies $1 \ge S_{\epsilon} \ge 0$. The entanglement entropy of the system can be defined as the von neuman entropy of 
either subsystem  is equal to $S_{\epsilon}$. In Fig. \ref{fig:entropy} (a), (b) and(d) $S_{\epsilon}$ is plotted 
along the color axis for different initial condition and different values of the detuning parameter $\delta$. By (\ref{EE}) $S_{\epsilon}$ is entirely determined by $|C_{2}(t)|^{2}$. Therefore the variation of $S_{\epsilon}$ is very similar to the corresponding variation of $|C_{2}(t)|^{2}$ given in Fig.   \ref{fig:Rabi}. 
 
For, $|\Psi_{SO} \rangle=|\uparrow,k_x+k_o\rangle $, the state
is a product state ( $C_{2}=1, C_{1}=0$), located at the north pole of the Bloch sphere in Fig. \ref{fig:BB} (a).
From the  Fig. \ref{fig:Rabi}(a) and (b) and Fig. \ref{fig:entropy} (a), (b)
for such case the entangleemnt entropy $S_{\epsilon}=0$. Whereas  the figures show that for maximally entangled state such as 
$|\Psi_{SO}\rangle=\frac{1}{\sqrt{2}}(|\uparrow,k_x+k_o\rangle+|\dna,k_x-k_o\rangle $  ($C_{1}=C_{2}=\frac{1}{\sqrt{2}}$) (indicated in the equitorial plane of the Bloch sphere in Fig. \ref{fig:BB} (a)), $S_{\epsilon}=1$, its maximum value. 

Another well known measure of  entanglement in a system is concurrence which is originally formulated  to measure entanglement in mixed quantum state. Later, Hill and Wootter\cite{Wootter1, Wootter2} introduce concurrence for bipartite pure state as 
$C=\sqrt{2(1-Tr\rho^2)}$
Non zero value of concurrence C confirmed the entanglement in system
where $\rho$ is reduce density matrix of system. For the pure state under consideration  $\rho=\rho_s=\rho_m$.
This straightforwardedly gives 
$C=2C_1(t)C_2(t)$ which is $\ge 0$ for a general state (\ref{ENT}). As expected $C$ also shows same periodic variation in time as $S_{\epsilon}$ and $|C_{2}(t)|^{2}$ in Fig. \ref{fig:entropy} (c).

\section{Viloation of Bell like Inequality}
Given the fact that the state (\ref{ENT})  is entangled in  spin and momentum, it is expected to violate 
Bell's Inequality \cite{Gisin}. 
The relevant inequality in this case is 
CHSH inequality\cite{CHSH}, which is a generalisation of BI and refer as Bell-like inequality. 
To set the frame work for testing BI we first define our basis states for different set of measurements which are obtained from the basis states defined in (\ref{basis}) by unitary transformtion. They are 
\beq \bs{u}^{\alpha}=R( \alpha) \bs{u} , \bs{v}^{\beta} = R(\beta) \bs{v} \nonumber \eeq 
with $R(\theta) = \begin{bmatrix} \cos \theta & \sin \theta \\ -\sin \theta & \cos \theta \end{bmatrix}$
and $\bs{u} = (|u_{1} \rangle, |u_{2} \rangle)^{T}, \bs{v}=(|v_{1} \rangle, |v_{2} \rangle)^{T}$.
Here $\alpha, \beta$ corresponds to the angle made by the the detectors that are measuring the mutually commutating observables, pseudopsin and momentum. 
In terms of these transformed basis states the CHSH inequality  becomes 
 \beq
{\ss}
=|E(\alpha,\beta)+E(\alpha',\beta)+E(\alpha',\beta')-E(\alpha,\beta')|\leq 2
 \label{CHSK} \eeq
 where each $E(\alpha,\beta)$ is the correlation between spin $|u_i^\alpha\rangle$ and momentum $|v_j^\beta\rangle$ degree of freedom of the ultra cold atom
and is given by 
\beq E(\alpha, \beta)= P_{11}(\alpha, \beta) -P_{12}(\alpha, \beta) -P_{21}(\alpha, \beta) + P_{22}(\alpha, \beta). \label{BellC} \eeq
Here $P_{ij}(\alpha,\beta)$ is the probability of getting the atom in the product state $|u_i^\alpha\rangle\otimes|v_j^\beta\rangle$ and are given by 
$P_{ij}(\alpha,\beta)=\langle\Psi_{SO}|\hat{P_i^\alpha}\hat{P_j^\beta}|\Psi_{SO}\rangle$.
Here $\hat{P}^{\alpha, \beta}_{i}$ s are projection operators and given by $\hat{P}^{\alpha}_{i}=  | u_{i}^{\alpha} \rangle \langle u_{i}^{\alpha}|$ and $P_{i}^{\beta} = | v_{i}^{\beta} \rangle \langle v_{i}^{\beta}|$. 
From (\ref{ENT}) and (\ref{BellC}) one therefore gets 
\beq
E(\alpha,\beta)=\cos(2\alpha)\cos(2\beta)+(C_1 C_2^* +C_1^* C_2)\sin(2\alpha)\sin(2\beta) \eeq
which can be used to calculate ${\ss}$ in (\ref{CHSK}) to verify BI for the current problem. 
It is well known that all choices of  $\alpha$ and $\beta$ do not satisfy this inequality \cite{Sakurai}. 
We made the choice 
$ \alpha=0,\quad \beta=\frac{\pi}{8},\quad \alpha'=\frac{\pi}{4}, \quad \beta'=\frac{3\pi}{8}$ for which 
${\ss}=\sqrt{2}(C_1 C_2^*+C_1^* C_2+1)$.
In Fig. \ref{fig:BB} (b) the above expression for the parameter ${\ss}$ is plotted as function of $\frac{\Omega}{E_{0}}$ for the initial conditions mentioned in Fig. \ref{fig:Rabi}(a) and (b). 
As expected from the preceeding discussion and the expression, ${\ss}$ also shows oscillation with time for a given $\frac{\Omega}{E_{0}}$.
As can be seen for a wide range values for $\frac{\Omega}{E_{0}}$ at different detuning $\delta$, the inequality is violated, namely ${\ss} > 2$. When the state (\ref{ENT}) is a maximally entangled, namnely 
$C_{1}=C_{2}=\frac{1}{\sqrt{2}}$, ${\ss}=2\sqrt{2}$ reaching the maximum bound of the Bell Inequality violation, that is  $2\sqrt{2}$ given by Cirelson \cite{Cirel}. 
On the otherhand when it is a simple product state ( for example, $C_{1}=0, C_{2}=1$), ${\ss}=\sqrt{2}$, the BI is obeyed. 

\begin{figure}
\centering
\includegraphics[width=1.05\columnwidth , height= 
0.45\columnwidth]{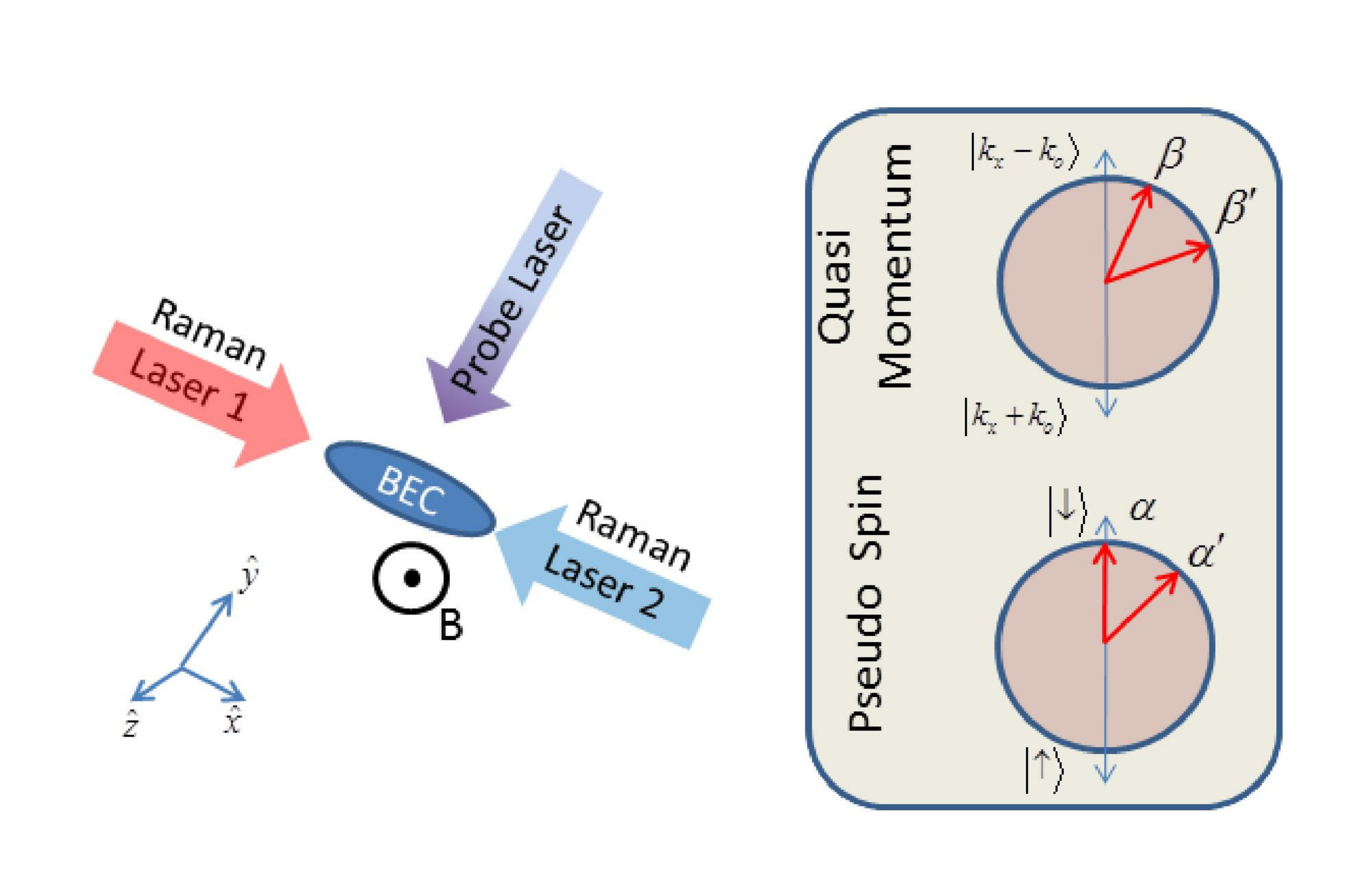} 
\caption{(Color online) Schematic diagram for the spin and momentm measurement for testing BI. In the circles 
we show schematically the transformed basis in the spin and momentum space.}
\label{fig:Bell_test}
\end{figure}

For experimental testing of BI we required a ensemble of such atoms. We can calculate the corresponding correlation $E(\alpha,\beta)$ through statistical measurement given as$E(\alpha,\beta)=\frac{N_{11}(\alpha,\beta)+N_{22}(\alpha,\beta)-N_{12}(\alpha,\beta)-N_{21}(\alpha,\beta)}{N_{11}(\alpha,\beta)+N_{22}(\alpha,\beta)+N_{12}(\alpha,\beta)+N_{21}(\alpha,\beta)}$ with $P_{ij}(\alpha, \beta)=\frac{N_{ij}(\alpha,\beta)}{\sum_{i,j=1}^{2}N_{ij}(\alpha, \beta)}$. 
Stern-Gerlach (SG)set up is a unique way of spin detection and was already used for measurement in case of ultracold spinorial condensate \cite{Fu}. Here we have to make spin measurement for two different settings $\alpha$ and $\alpha'$ of SG set-up which could be, for example,  
 the angle from +z axes if we consider the pseudo spin in +z direction. 
A schematic of such set up is depicted in Fig. \ref{fig:Bell_test}. One can further do momentum spectroscopy 
on the ultra cold atomic cloud to extract the number of atoms collapsing in the state $|v_{j}^{\beta} \rangle$
with a given spin -orientation.From such joint measurements the parameter ${\ss}$ can be determined. 

Our work can lead to interesting possibilities which has both fundamental and applied interest.  
It can be starightforwardedly extended to  experimentally realized ultra cold fermions \cite{FermiSO1, FermiSO2} with spin-orbit coupling. Coupling such spin-orbit entangled atom with each other or with 
photons it is also possible to create systems with different type of hybrid entanglement \cite{Ma}. 
Since each atom is entangled in spin and momentum degrees of freedom and shows Rabi oscillation, a periodic array of such atoms in an optical lattice can implement SWAP operation in large scale quantum computation \cite{review2}. To summarize 
this work demonstarted exciting possibility of testing certain fundamental issues of quantum mechanics such as 
quantum non-contextuality and BI violation
with synthetically spin-orbit coupled ultra cold bosons  and potent with extremely rich possibilities.
We thank Emroj Hossain for helpful discussion during the early stage of the project. RK is partially supported by a DST INSPIRE scholarship.

\end{document}